\documentclass[aps,prb,twocolumn,showpacs,superscriptaddress,longbibliography]{revtex4-2}
\usepackage{amsfonts}
\usepackage{amsmath}
\usepackage{braket}
\usepackage{graphicx}
\usepackage{bm}
\usepackage{amssymb}
\usepackage{dcolumn}
\usepackage{color}
\usepackage{soul}
\usepackage{multirow}
\usepackage{booktabs}
\usepackage{diagbox}
\usepackage{verbatim}
\usepackage[colorlinks,
    linkcolor=blue,
    anchorcolor=blue,
    citecolor=blue,
    urlcolor=blue,
]{hyperref}

\begin{document}

    \title{Ferroelectric control of bipolar magnetic semiconductor with room Curie temperature}
	
	\author{Jia-Wen Li}
	\affiliation{Kavli Institute for Theoretical Sciences, University of Chinese Academy of Sciences, Beijing 100049, China}
	
	\author{Gang Su}
	\email{gsu@ucas.ac.cn}
	\affiliation{Kavli Institute for Theoretical Sciences, University of Chinese Academy of Sciences, Beijing 100049, China}
	\affiliation{Physical Science Laboratory, Huairou National Comprehensive Science Center, Beijing 101400, China}
	\affiliation{Institute of Theoretical Physics, Chinese Academy of Sciences, Beijing 100190, China}
	\affiliation{School of Physical Sciences, University of Chinese Academy of Sciences, Beijing 100049, China}
	
	\author{Bo Gu}
	\email{gubo@ucas.ac.cn}
	\affiliation{Kavli Institute for Theoretical Sciences, University of Chinese Academy of Sciences, Beijing 100049, China}
	\affiliation{Physical Science Laboratory, Huairou National Comprehensive Science Center, Beijing 101400, China}

	    \begin{abstract}
	    	
	    	The development of room-temperature tunable magnetic semiconductors is crucial for the advancement of low-power, high-performance information technologies.
	    	Using density functional theory calculations, we propose a series of two-dimensional magnetic semiconductors with critical temperature above room temperature, including three ferromagnetic and two antiferromagnetic semiconductors.
	    	Their stability is confirmed through phonon spectra, molecular dynamics simulations, and formation energy calculations.
	    	In particular, we demonstrate a ferromagnetic bipolar magnetic semiconductor (BMS), Cr$_2$NiSe$_4$, formed via Ni intercalation into bilayer CrSe$_2$, which exhibits a 0.40 eV band gap and a Curie temperature of 352 K.
	    	Nonvolatile carrier spin polarization control in Cr$_2$NiSe$_4$ is achieved by switching the ferroelectric polarization of an Al$_2$Se$_3$ substrate.
	    	Switching the ferroelectric state of monolayer Al$_2$Se$_3$ induces a BMS-to-half-metal transition.
	    	Reversing the polarization of bilayer Al$_2$Se$_3$ yields a half-metallic Cr$_2$NiSe$_4$ with fully opposite carrier spin polarization.
	    	Furthermore, we propose a multiferroic nonvolatile memory design: write operations are controlled by the ferroelectric polarization state of bilayer Al$_2$Se$_3$, while read operations rely on detecting the distinct carrier spin polarizations of Cr$_2$NiSe$_4$.
	    	Our work reports a two dimensional BMS with Curie temperature above room temperature and presents a feasible strategy for its nonvolatile electrical control.
    \end{abstract}
    \maketitle


    \section{Introduction}

    The pursuit of information technologies beyond the confines of Moore's Law has catalyzed extensive research into spintronic devices that promise enhanced speed, smaller footprints, and lower energy consumption \cite{Shalf2020,Hu2015,Wang2021}.
    Due to the interesting properties, magnetic semiconductors have many promising applications in spintronics \cite{Dietl2010,Ohno2010,Sato2010,Jungwirth2006,Dietl2014,Fiederling1999,Ohno2000,Mitra2001,Song2018,Li2019,Gorbenko2007,Goel2023,Cinchetti2008}.
    A persistent bottleneck, however, is that most experimentally realized ferromagnetic semiconductors exhibit a Curie temperature ($T{\rm_C}$) far below room temperature, which largely limits their applications.
    In 2017, the successful synthesis of two-dimensional (2D) van der Waals ferromagnetic semiconductors CrI$_3$ \cite{Huang2017} and Cr$_2$Ge$_2$Te$_6$ \cite{Gong2017} in experiments has attracted extensive attention to 2D ferromagnetic semiconductors.
    According to Mermin-Wagner theorem \cite{Mermin1966}, the magnetic anisotropy is essential to produce long-range magnetic order in 2D systems.
    Recently, with great progress of 2D magnetic materials in experiments, more 2D ferromagnetic materials have been obtained, where some are ferromagnetic semiconductors with $T\rm_C$ far below room temperature  \cite{Chu2019,Cai2019,Zhang2019,Achinuq2021,Lee2021}, and some
    are ferromagnetic metals with high $T\rm_C$ above room temperature  \cite{Zhang2021,Xian2022,Wang2024,Chua2021,Li2022b,Zhang2019a,Deng2018,Fei2018,Seo2020,Zhang2022,Chen2024,O’hara2018,Xiao2022,Yao2024,Bonilla2018,Meng2021}.
    In addition, many ferromagnetic semiconductors with $T{\rm_C}$ above room temperature have been predicted based on theoretical calculations \cite{Huang2019a,Jiang2018a,Huang2018a,Li2023b,Li2024a,Li2024,Li2025}, while their synthesis remains a challenge.
    
    The inherent structural uniqueness of two-dimensional (2D) materials renders their properties exceptionally sensitive to external manipulation, offering pathways for precise control.
    Prominent strategies to achieve such control include electric field application \cite{Deng2018,Jiang2018,Zhao2021}, chemical doping \cite{Hao2019,Fang2022,Zhao2022a}, strain engineering \cite{Huang2019,Li2020,Zhang2021}, and atomic intercalation \cite{Pathirage2023,Zhou2021a,Rajapakse2021,Zhao2024,Zhou2022,Peng2023,Zhang2017,Liu2020,Wang2024a}.
    Among these, atomic intercalation is a particularly powerful and widely utilized method.
    This technique is facilitated by the layered architecture of van der Waals (vdW) materials, which allows for the insertion of guest atoms into the interstitial vdW gaps, thereby altering the host's structure and properties \cite{Rajapakse2021}.
    The versatility of this approach is well-documented; for instance, Cr intercalation in 1T-CrTe$_2$ \cite{Zhang2021} yields a family of derivative compounds such as Cr$_3$Te$_4$ \cite{Wang2022}, Cr$_2$Te$_3$ \cite{Wen2020}, and Cr$_5$Te$_8$ \cite{Chen2021,Tang2022}, among others \cite{Liu2022a,Tan2023}.
    Similarly, Cr and Mn atoms can be periodically inserted into the vdW gaps of VSe$_2$ \cite{Pathirage2023}.
    The intercalation process can be remarkably efficient, with Cu atoms spontaneously intercalate into layered group IV/V MX$_2$ TMDs (M = Ti, V, Ta, Nb; X = S, Se) up to 1.2 per formula cell \cite{Liu2020}, and Bi$_2$Se$_3$ accommodates Cu up to 6.7 per formula cell \cite{Zhang2017}.
    Furthermore, intercalation can profoundly engineer functionality, such as inducing a ferromagnetic-to-antiferromagnetic transition in CrSe$_2$ (forming CuCrSe$_2$) \cite{Peng2023} or triggering room-temperature ferromagnetism in otherwise non-magnetic WO$_3$ \cite{Zhao2024}.
    The insertion of 1D VS chains into 2D VS$_2$ also creates hetero-dimensional VS$_2$-VS superlattices \cite{Zhou2022}.

    The bipolar magnetic semiconductor (BMS) is a special magnetic semiconductor, whose valence band maximum (VBM) and conduction band minimum (CBM) possess opposite spin polarization direction \cite{Li2012,Li2022c}.
    This distinct band structure allows an applied gate voltage to shift the Fermi level, inducing a transition to a half-metallic (HM) state with fully spin-polarized carriers of either spin-up or spin-down character.
    Consequently, BMSs are highly promising candidates for advanced spintronic applications, including bipolar field-effect spin filters and spin valves \cite{Li2022c}, platforms for the spin Seebeck effect \cite{Ding2020}, bipolar pseudospin systems \cite{Jung2020}, and even tools for detecting entangled electrons from superconductors \cite{Li2013}.
    Despite these compelling prospects, the exploration of BMSs has been limited, with only a handful of candidates theoretically proposed in recent years \cite{Guo2024,Li2018a,Cheng2018,Sheng2021,Liu2024,Lv2021,Wang2022a,Wang2022b,Li2023,Zhang2024a}.
    The translation of BMS into practical technologies is principally impeded by two key challenges.
    Firstly, a high $T\rm_C$ exceeding room temperature is essential for practical devices.
    Secondly, the spin polarization of carriers in BMS is typically controlled by volatile electrical gating, requiring continuous application of an electric field and resulting in substantial energy consumption \cite{Zhao2019,Li2023}.
    Therefore, achieving nonvolatile control over a BMS with $T\rm_C$ above room temperature represents a critical step toward realizing their technological promise.

    In this work, we address these outstanding challenges by systematically designing a 2D room-temperature BMS and demonstrating its nonvolatile control using a ferroelectric gate.
    Our strategy begins with first-principles predictions based on transition-metal intercalation into bilayer CrS$_2$ and CrSe$_2$.
    We predict five magnetic semiconductors operable at room temperature, including three ferromagnetic (FM) and two antiferromagnetic (AFM), whose structural stability was confirmed via phonon spectra, molecular dynamics, and formation energy calculations.
    In particular, monolayer Cr$_2$NiSe$_4$, formed by Ni intercalation into bilayer CrSe$_2$, is revealed to be a ferromagnetic BMS with a substantial 0.40 eV band gap and a $T{\rm_C}$ of 352 K.
    Inspired by the ferroelectric materials with spontaneous electric polarization, we propose a general strategy to achieve nonvolatile spin polarization control in BMS using a ferroelectric gate.
    Ferroelectric monolayer (ML-) or bilayer (BL-) Al$_2$Se$_3$ are chosen as the ferroelectric gate \cite{Ding2017,Li2023,Yuan2025}.
    Our calculations show that different ferroelectric polarization directions yield distinct band alignments, which in turn induces the injection of either electrons or holes into the BMS layer. These distinct carrier injections result in distinct Fermi energy shifts, leading to half-metallic states with distinct carrier spin polarizations. 
    Specifically, in the Cr$_2$NiSe$_4$/ML-Al$_2$Se$_3$ heterostructure, reversing the ferroelectric polarization of Al$_2$Se$_3$ induces a transition in Cr$_2$NiSe$_4$ between a half-metallic and a semiconducting state.
    In the Cr$_2$NiSe$_4$/BL-Al$_2$Se$_3$ heterostructure, different ferroelectric polarization directions of BL-Al$_2$Se$_3$ result in Cr$_2$NiSe$_4$ being a half metal with opposite carrier spin polarization. 
    Based on these findings, nonvolatile multiferroic memories were proposed using Cr$_2$NiSe$_4$/BL-Al$_2$Se$_3$ heterostructures, where data writing is controlled by the ferroelectric polarization of 2D Al$_2$Se$_3$ via an electric signal, and data reading exploits the distinct carrier spin polarization of Cr$_2$NiSe$_4$.
    This device offers the advantages of convenient nondestructive reading and writing, high-temperature operation, and low energy consumption for data storage.
    Our work reports a 2D BMS Cr$_2$NiSe$_4$ with $T\rm_{C}$ above room temperature and present a feasible strategy for its nonvolatile electrical control.

	\section{Results and Discussion}
	
    \subsection{Crystal structure, stability and electric structure of Cr$_2$NiSe$_4$}

\begin{figure}[hbpt]
		\centering
		\includegraphics[width=0.95\columnwidth]{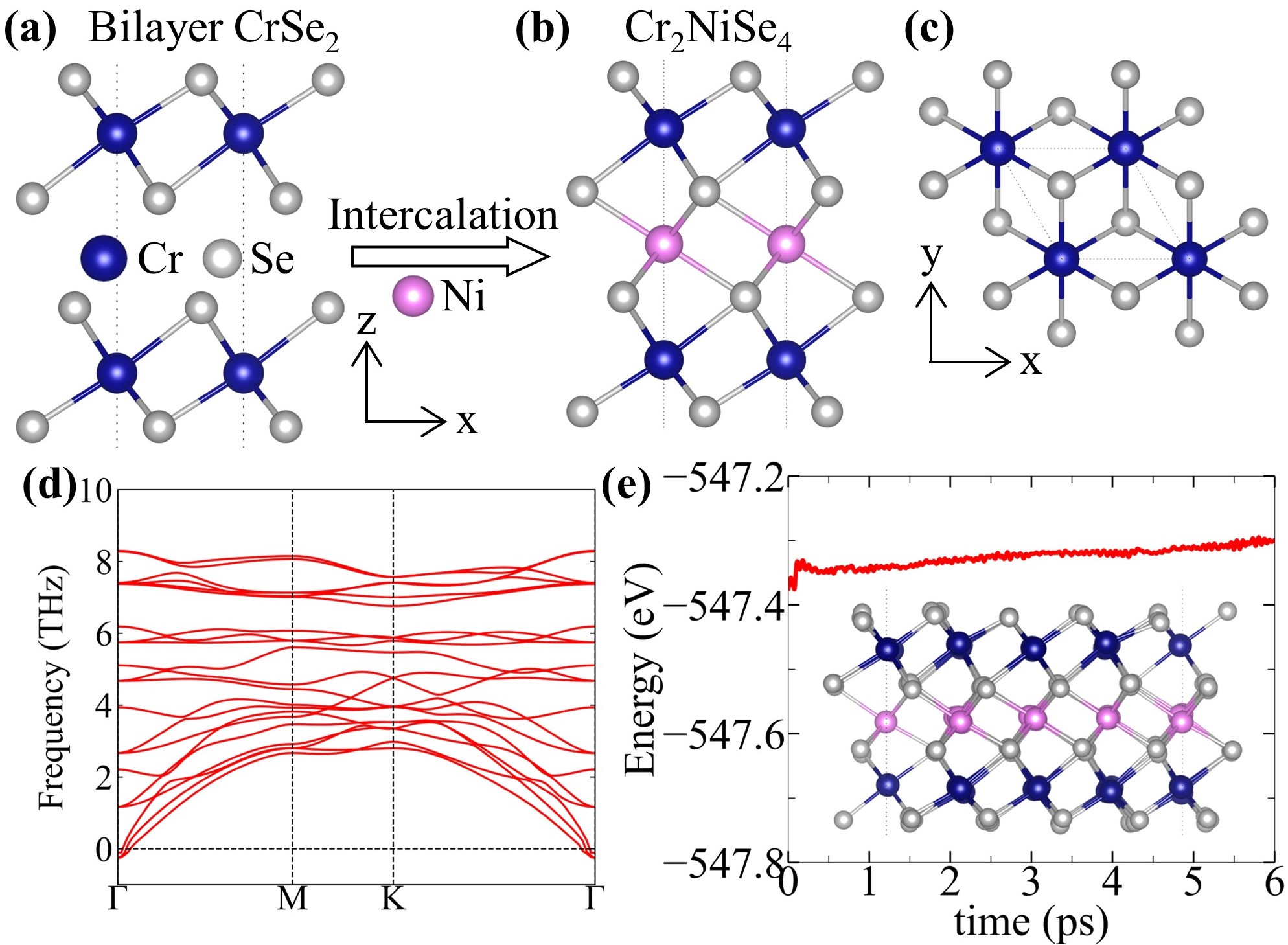}\\
		\caption{
			Structure and stability analysis of monolayer Cr$_2$NiSe$_4$. 
			(a) Side view of bilayer 1T-CrSe$_2$.
			(b) Top and (c) side views of the crystal structure of monolayer Cr$_2$NiSe$_4$, formed by Ni intercalation.
			(d) Phonon spectrum of monolayer Cr$_2$NiSe$_4$, confirming its dynamical stability.
			(e) Total energy evolution from an ab initio molecular dynamics simulation at 300 K for 6 ps, confirming its thermodynamic stability.
			The inset shows the final atomic structure after the simulation.
		}\label{CS-Ni1}
	\end{figure}

	
	The crystal structure of bilayer hexagonal 1T-CrSe$_2$ is shown in Fig. \ref{CS-Ni1}(a), with space group P$\overline{3}$m1 (164).
	The calculated in-plane lattice constants are $a_0=b_0=3.56$ \AA, agree with the experimental value of 3.63~\AA~\cite{Li2021}.
	The interlayer distance is 2.55 \AA.
	According to our DFT results, bilayer CrSe$_2$ is an AFM metal with intralayer FM coupling and interlayer AFM coupling (A-AFM), agree with previous reports \cite{Wu2022,Wang2020a}.
	The detailed results are provided in the Supplementary Material \cite{SM}.
	
	Monolayer Cr$_2$NiSe$_4$ was obtained by intercalation Ni into the vdW gap of bilayer CrSe$_2$.
	The crystal structure of monolayer Cr$_2$NiSe$_4$ is shown in Figs. \ref{CS-Ni1}(b) and \ref{CS-Ni1}(c).
	The in-plane lattice constants increased by 3.6\% to 3.69 \AA~compared to BL-CrSe$_2$. 
	The interlayer distance increased by 8\% to 2.76 \AA.
	The phonon spectrum of monolayer Cr$_2$NiSe$_4$, as shown in Fig. \ref{CS-Ni1}(d), exhibits negligible negative frequencies, indicating its stability.
	In addition, we performed molecular dynamics simulations of Cr$_2$NiSe$_4$ at 300 K using the NVT ensemble (constant temperature and volume) for 6 ps.
	The results show it is thermodynamically stable (Fig. \ref{CS-Ni1}(e)).
	To further analyze the stability of structures, the formation energy $E\rm_{formation}$ of Cr$_2$NiSe$_4$ is calculated by $E{\rm_{formation}}=
	\left(E{\rm_{Cr_2NiSe_4}}-2E{\rm_{Cr}}-E{\rm_{Ni}}-4E{\rm_{Se}}\right)/7$, where $E{\rm_{Cr_2NiSe_4}}$ is the energy of Cr$_2$NiSe$_4$.
	$E\rm_{Cr}$, $E\rm_{Ni}$ and $E\rm_{Se}$ are energies per atom for bulks of Cr, Ni and Se with the lowest energy, respectively.
	The formation energy $E\rm_{formation}$ of Cr$_2$NiSe$_4$ is -1.13 eV/atom.
	The negative sign indicates that the energy of these atoms in the compound is lower than that in the case of simple substances.
	In addition, the formation energy is lower than that of -1.09 eV/atom in bilayer CrSe$_2$, indicating the structural stability.
	
	\begin{figure}[hbpt]
		\centering
		\includegraphics[width=0.99\columnwidth]{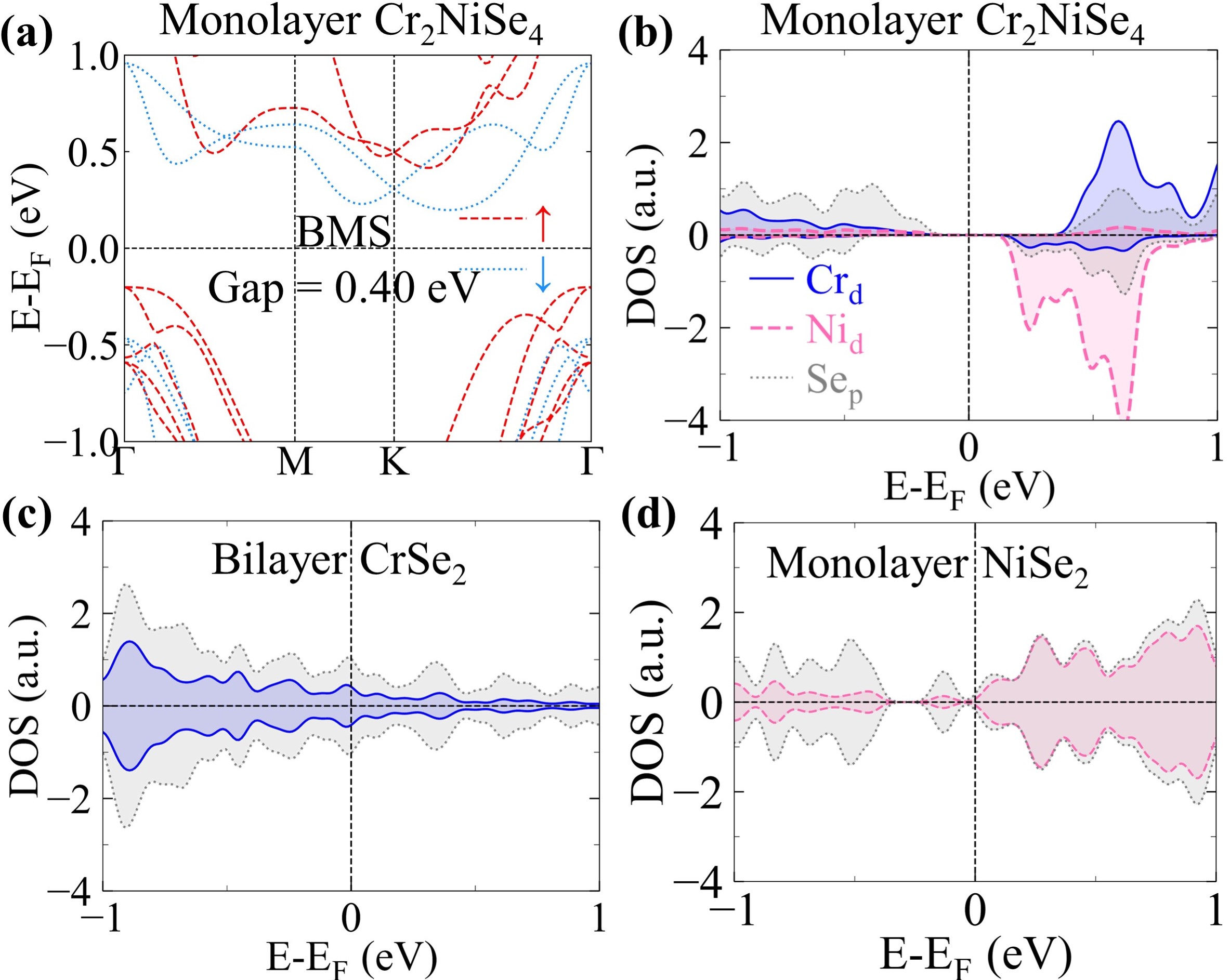}\\
		\caption{
			Electric structures of monolayer BMS Cr$_2$NiSe$_4$, bilayer CrSe$_2$ and monolayer NiSe$_2$.
			(a) DFT results of spin polarized band structure of Cr$_2$NiSe$_4$, showing BMS properties.
			(b-d) Spin polarized partial density of states of (b) monolayer Cr$_2$NiSe$_4$, (c) bilayer CrSe$_2$ and (d) monolayer NiSe$_2$ around the Fermi energy.
		}\label{M2S}
	\end{figure}

	The DFT-calculated band structure of Cr$_2$NiSe$_4$ is displayed in Fig. \ref{M2S}(a), exhibiting a band gap of 0.40 eV.
	Cr$_2$NiSe$_4$ is a BMS with a spin-up polarized VBM and a spin-down polarized CBM.
	Its BMS behavior persists even with HSE hybrid functional calculations \cite{Heyd2003}, as shown in Supplementary Material \cite{SM}.

	To understand the metal-semiconductor transition from bilayer CrSe$_2$ to monolayer Cr$_2$NiSe$_4$ via Ni intercalation, we calculated the partial density of states (PDOS) for monolayer Cr$_2$NiSe$_4$, bilayer CrSe$_2$, and monolayer NiSe$_2$.
	These are shown in Figs.~\ref{M2S}(b), \ref{M2S}(c), and \ref{M2S}(d), respectively.
	For bilayer CrSe$_2$, the presence of non-zero PDOS at the Fermi level ($E_F$) coupled with the inherent symmetry of its spin-up and spin-down PDOS conclusively confirms its AFM metallic character.
	These electronic states are predominantly derived from the hybridized contributions of Cr $d$-orbitals and Se $p$-orbitals.
	Monolayer NiSe$_2$ displays AFM PDOS with small band gap (Fig.~\ref{M2S}(d)), formed primarily by Ni $d$ and Se $p$ orbitals, consistent with previous work \cite{Bravo2023}.
	As shown in Fig.~\ref{M2S}(b), the intercalation of Ni atoms introduces Ni $d$-orbitals near $E_F$.
	This structural perturbation and the introduction of Ni-derived states lead to a substantial rearrangement of the energy bands in Cr$_2$NiSe$_4$.
	Notably, results of the PDOS reveals that the CBM in Cr$_2$NiSe$_4$ becomes dominated by Ni $d$-orbitals, while the VBM is primarily composed of Se $p$-orbitals.
	This electronic re-hybridization effectively depletes the DOS at the Fermi level, culminating in the formation of a 0.40~eV band gap and thereby driving the observed metal-to-semiconductor transition.

	\subsection{Magnetic properties of Cr$_2$NiSe$_4$}
	
\begin{figure}[hpbt]
		\centering
		\includegraphics[width=0.96\columnwidth]{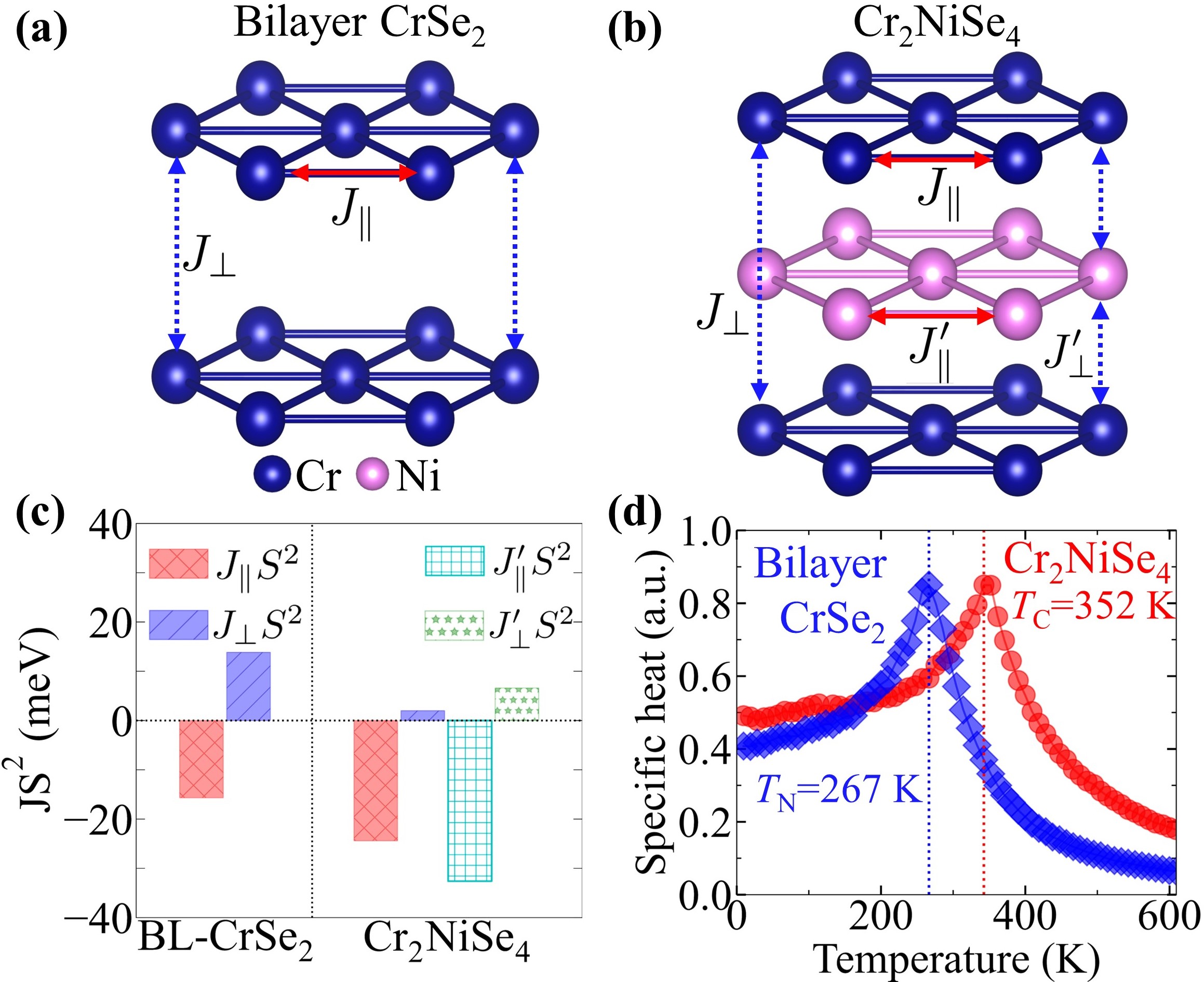}\\
		\caption{
			Magnetic properties of bilayer CrSe$_2$ and monolayer Cr$_2$NiSe$_4$.
			(a, b) Schematics defining the magnetic coupling constants for (a) bilayer CrSe$_2$ and (b) monolayer Cr$_2$NiSe$_4$.
			(c) Calculated values of the corresponding exchange coupling constants ($JS^2$) for both structures.
			(d) Temperature-dependent specific heat calculated from Monte Carlo simulations using the 2D Heisenberg model in Eq. \eqref{eq:2D}.
			The peaks indicate a Neel temperature ($T\rm_N$) of 267 K for bilayer CrSe$_2$ and a Curie temperature ($T\rm_C$) of 352 K for monolayer Cr$_2$NiSe$_4$.
		}\label{Mag}
	\end{figure}

	To analyze the magnetic properties of Cr$_2$NiSe$_4$, we model the system as a three-layer hexagonal magnetic lattice composed of Cr and Ni atoms, as schematically shown in Fig. \ref{Mag}(b).
	The magnetic interactions are described by a 2D Heisenberg-type Hamiltonian:
	\begin{align}
		H=&~
		J_\parallel\sum\limits_{<i,j>}^{intra.}\vec{S_i}\cdot \vec{S_j}
		+J_\perp\sum\limits_{<i,j>}^{inter.}\vec{S_i} \cdot \vec{S_j}
		+
		J'_\parallel\sum\limits_{<i,j>}^{intra.}\vec{S_i}'\cdot \vec{S_j}'
		\nonumber\\&+
		J'_\perp\sum\limits_{<i,j>}^{inter.}\vec{S_i} \cdot \vec{S_j}'
		+A\sum\limits_{i}S_{iz}^2,
		\label{eq:2D}
	\end{align}
	where $\vec{S_i}$ and $\vec{S_j}'$ are the spin operators for Cr atoms at site $i$ and Ni atoms at site $j$, respectively.
	The Hamiltonian includes two intralayer exchange coupling constants, $J_\parallel$ (Cr-Cr) and $J'_\parallel$ (Ni-Ni), and two interlayer coupling constants, $J_\perp$ (next-nearest Cr-Cr) and $J'_\perp$ (nearest Cr-Ni).
	The parameter $A$ represents the single-ion magnetic anisotropy, defined as $AS^2=(E_{\perp}-E_{\parallel})/N_{\rm{mag}}$, where $E_{\perp}$ and $E_{\parallel}$ are the total energies from DFT calculations for out-of-plane and in-plane magnetization, respectively, and $N_{\rm{Mag}}$ is the number of magnetic atoms in the unit cell.
	Our calculation yields $AS^2 = 0.18$ meV per magnetic atom, and its positive sign indicates a preference for in-plane magnetization.

	\begin{table*}[hpbt]
		\setlength{\tabcolsep}{1.5mm}
		\caption{
			Family of room temperature magnetic
			semiconductors Cr$_2$CdS$_4$ and Cr$_2$XSe$_4$ (X = Zn, Cd, Mn, Ni). 
			The band gap, formation energy $E\rm_{formation}$, coupling constants, MAE, Curie temperature $T\rm_C$ or Neel temperature $T\rm_N$ for the predicted magnetic semiconductors.
			The results are obtained by DFT calculations and Monte Carlo simulations.
			A-AFM represent intralayer FM coupling and interlayer AFM coupling.
			$E\rm_{formation}$ of bilayer-CrS$_2$ and bilayer-CrSe$_2$ are -1.09 eV/atom and -1.46 eV/atom, respectively.
		}{\scalebox{1}{
				\begin{tabular}{c|c|c|c|c|c|c|c|c|c|c}
					\hline\hline
					\multicolumn{2}{c|}{Magnetic semiconductors} &
					\multicolumn{9}{c}{Properties}
					\\
					\hline
					\makebox[0.10 \textwidth][c]
					{\multirow{2}{*}{Host}} &
					\makebox[0.10 \textwidth][c]
					{\multirow{2}{*}{Intercalation}} &
					\makebox[0.08 \textwidth][c]
					{\multirow{2}{*}{Gap (eV)}}&
					\makebox[0.08 \textwidth][c]{$E{\rm_{formation}}$}&
					\multicolumn{4}{c|}{$JS^2$ (meV)} &
					\makebox[0.06 \textwidth][c]
					{MAE} &
					\makebox[0.08 \textwidth][c]{Ground}&
					\makebox[0.07 \textwidth][c]
					{$T\rm_C$ or $T\rm_N$ }
					\\
					\cline{5-8}
					&&
					&
					(eV/N)&
					$J_\parallel S^2$&
					$J_\perp S^2$&
					$J_\parallel' S^2$&
					$J_\perp' S^2$&
					(meV)
					&State
					& (K)
					\\
					\hline
					\multirow{1}{*}{Bilayer-CrS$_2$}&
					Cd & 0.98 & -1.23 &-23.67  &-0.12 & 0& 0 & -0.11 & FM& 319 \\
					\hline
					\multirow{4}{*}{Bilayer-CrSe$_2$}&
					Zn & 0.96 & -1.15
					&-24.61 & 1.05 & 0& 0  &0.14 & A-AFM& 324 ($T\rm_N$) \\
					& Cd & 0.81 & -1.13 
					&-26.24 & 1.44 & 0 & 0 & 0.06  & A-AFM& 352 ($T\rm_N$)\\\cline{2-11}
					& Mn& 0.48 & -1.34
					& -25.3 & 1.04&  -27.01 & 4.29 &0.18 &FM& 371 \\
					& Ni& 0.40 (BMS)& -1.13 
					& -24.4 &2.00& -30.59 & 6.57 & 0.18 &FM& 352 \\
					\hline\hline
				\end{tabular}
		}}
		\label{tab:res}
	\end{table*}

	Our DFT calculations reveal that Cr$_2$NiSe$_4$ possesses a FM ground state.
	The calculated exchange coupling constants are $J_\parallel S^2 = -24.4$ meV, $J'_\parallel S^2 = -30.6$ meV, $J_\perp S^2 = 2.0$ meV, and $J'_\perp S^2 = 6.6$ meV, where negative and positive values denote FM and AFM coupling, respectively.
	By the Monte Carlo simulation based on the 2D Heisenberg model in Eq. \eqref{eq:2D}, the temperature dependent Specific heat of monolayer Cr$_2$NiSe$_4$ were calculated, as shown in Fig. \ref{Mag}(d), giving a $T\rm_C$ = 352 K.
	Detailed calculations of coupling constants are provided in the Supplementary Material \cite{SM}.
	Thus, a stable room-temperature BMS Cr$_2$NiSe$_4$ was obtained through Ni intercalation.

	To understand the AFM to FM transition by Ni intercalation, we compare these results with bilayer CrSe$_2$.
	Due to its bilayer structure, bilayer CrSe$_2$ only has non-zero $J_\parallel S^2$ and $J_\perp S^2$, as shown in Fig. \ref{Mag}(a), which were calculated as -15.67 meV and 13.85 meV, respectively.
	These coupling constants confirm bilayer CrSe$_2$ as A-type antiferromagnetic (A-AFM), characterized by intralayer ferromagnetic and interlayer antiferromagnetic coupling.
	Monte Carlo simulation give a $T\rm_N$ of 267 K, as shown in Fig. \ref{Mag}(d).
	The transition from an AFM state in bilayer CrSe$_2$ to FM state in Cr$_2$NiSe$_4$ is primarily attributed to the magnetic couplings induced by Ni layer.
	Upon intercalation, while the intralayer Cr-Cr coupling ($J_\parallel S^2$) remains ferromagnetic and even strengthens, the direct interlayer Cr-Cr AFM coupling ($J_\perp S^2$) is considerably weakened due to the increased layer separation.
	Crucially, the Ni layer introduces a strong intralayer Ni-Ni FM coupling ($J_\parallel' S^2$) and a significant interlayer Cr-Ni AFM coupling ($J_\perp' S^2$), which is much stronger than the residual Cr-Cr interlayer coupling (Fig. \ref{Mag}(c)).
	The competition between these interlayer coupling constants among these three FM layers ($J_\perp~vs~J_\perp'$) leads to the robust FM state in monolayer Cr$_2$NiSe$_4$.
	
	\subsection{
	Family of room temperature magnetic semiconductors Cr$_2$CdS$_4$ and Cr$_2$XSe$_4$ (X = Zn, Cd, Mn, Ni)}
		
	Intercalation of transition metal atoms into bilayer CrS$_2$ and CrSe$_2$ has yielded a family of stable high-$T\rm_C$ magnetic semiconductors, as listed in Table \ref{tab:res}.
	Five 2D room temperature magnetic semiconductors have been predicted, including 3 FM and 2 AFM semiconductors.
	Their stability is supported by phonon spectra exhibiting negligible negative frequencies, molecular dynamics simulations and negative formation energies.
	Their detailed properties are given in the Supplementary Material \cite{SM}.
	The discussion of different intercalation sites is provided in the Supplementary Material \cite{SM}, demonstrating that the cation site is the most stable intercalation position within the van der Wall gap.
	
	The intercalation of 3d, 4d, and 5d transition metal atoms, as well as alkali metals Li, Na, and K, into bulk CrX$_2$ (X = S, Se, and Te) was investigated.
	Some high $T\rm_N$ temperature AFM semiconductors with stable phonon spectrum were obtained, such as CrNaTe$_2$ with band gap of 0.21 eV and $T\rm_N$ of 210 K, and CrKTe$_2$ with band gap of 0.25 eV and $T\rm_N$ of 200 K.
	Their details are given in the Supplemental Material \cite{SM}.

	\subsection{Cr$_2$NiSe$_4$ at ferroelectric gate monolayer Al$_2$Se$_3$}

	\begin{figure*}[hbpt]
		\centering
		\includegraphics[width=1.83\columnwidth]{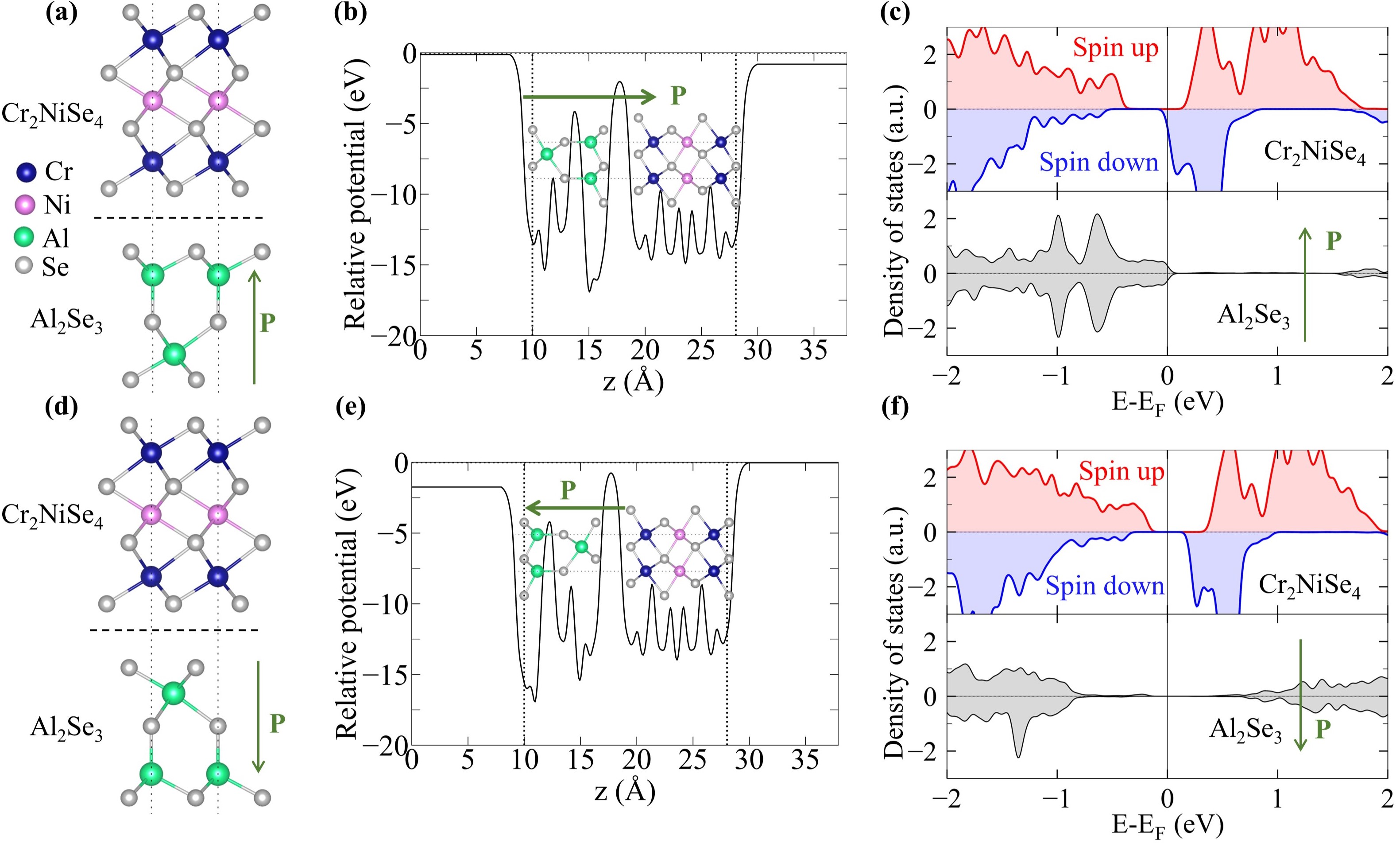}\\
		\caption{
			Cr$_2$NiSe$_4$/Al$_2$Se$_3$ heterostructures.
			(a) Side view, (b) local potential, and (c) layer resolved partial density of states of Cr$_2$NiSe$_4$/Al$_2$Se$_3$(P$\uparrow$).
			(d)-(f) The counterparts of Cr$_2$NiSe$_4$/Al$_2$Se$_3$(P$\downarrow$).
		}\label{AC}
	\end{figure*}

	The unique band structure of BMS allows for 100\% distinct spin-polarized carriers upon electron or hole doping.
	Cr$_2$NiSe$_4$ exhibits a half-metallic state with fully spin-polarized carriers when it loses or gains an electron. 
	This behavior was confirmed in our results of Cr$_2$NiSe$_4$ with charge doping, as given in the Supplementary Material \cite{SM}.
	However, the conventional method for controlling carrier spin polarization in a BMS relies on volatile electrical gating, which requires a continuous electric field and results in high energy consumption.
	
	Inspired by the ferroelectric materials with spontaneous electric polarization, nonvolatile spin polarization control in BMS with ferroelectric gate is possible \cite{Garcia2010,Kang2024,Li2023}.
	Ferroelectric materials exhibit nonvolatile polarization, maintaining their state without power consumption.
	Ferroelectric control offers lower energy consumption, smaller device size, and enhanced suitability for integrated applications in large-scale devices compared to gate voltage control.
	
	To implement this strategy for Cr$_2$NiSe$_4$, we employ monolayer and bilayer Al$_2$Se$_3$ as the ferroelectric gate.
	This material is a known 2D out-of-plane ferroelectric that has been successfully used in previous heterostructure device proposals \cite{Ding2017,Li2023,Yuan2025}.
	In our calculation, Al$_2$Se$_3$ show lattice constants of $a_1=b_1=3.79$~\AA, agree with the previous results of 3.79~\AA~\cite{Ding2017}.
	The lattice of Al$_2$Se$_3$ matches that of monolayer Cr$_2$NiSe$_4$ well with a mismatch~$\simeq$~2.6\%.
	The mismatch is obtained by $|a_0-a_1|/(a_0+a_1)\times100\%$, where $a_0$ and $a_1$ are the lattice constants of Cr$_2$NiSe$_4$ and Al$_2$Se$_3$, respectively.

	To demonstrate the proposed ferroelectric control, we constructed heterostructures of monolayer Cr$_2$NiSe$_4$ on monolayer Al$_2$Se$_3$.
	Two configurations were constructed, with the Al$_2$Se$_3$ pointing towards the Cr$_2$NiSe$_4$ layer (P$\uparrow$) and away from it (P$\downarrow$), as depicted in Figs. \ref{AC}(a) and \ref{AC}(d), respectively.
	The analysis is based on the most stable stacking configurations, as detailed in the Supplementary Material \cite{SM}.
	The plane averaged electrostatic potential along the $z$ direction for heterostructures is depicted in Fig. \ref{AC}(b) and \ref{AC}(e), respectively.
	The local electrostatic potential reduces/increases from bottom of Al$_2$Se$_3$(P$\uparrow$/P$\downarrow$) to Cr$_2$NiSe$_4$, showing the strong local polarization.

	\begin{figure*}[hbpt]
		\centering
		\includegraphics[width=1.83\columnwidth]{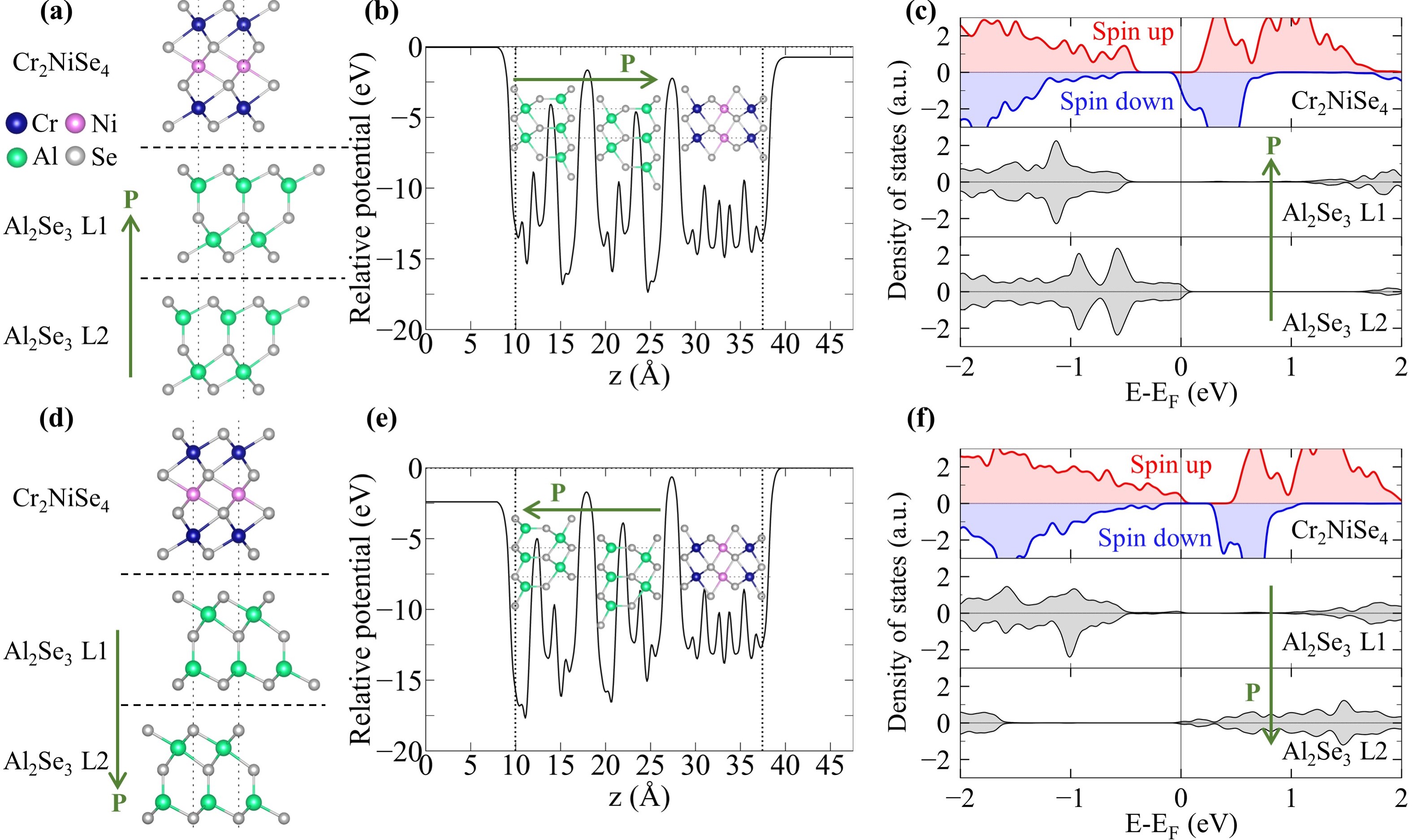}\\
		\caption{
			Cr$_2$NiSe$_4$/BL-Al$_2$Se$_3$ heterostructures.
			(a) Side view, (b) local potential, and (c) layer resolved partial density of states of Cr$_2$NiSe$_4$/BL-Al$_2$Se$_3$(P$\uparrow$).
			(d)-(f) The counterparts of Cr$_2$NiSe$_4$/Al$_2$Se$_3$(P$\downarrow$).
		}\label{A2C}
	\end{figure*}
	
	\begin{figure}[hbpt]
		\centering
		\includegraphics[width=0.96\columnwidth]{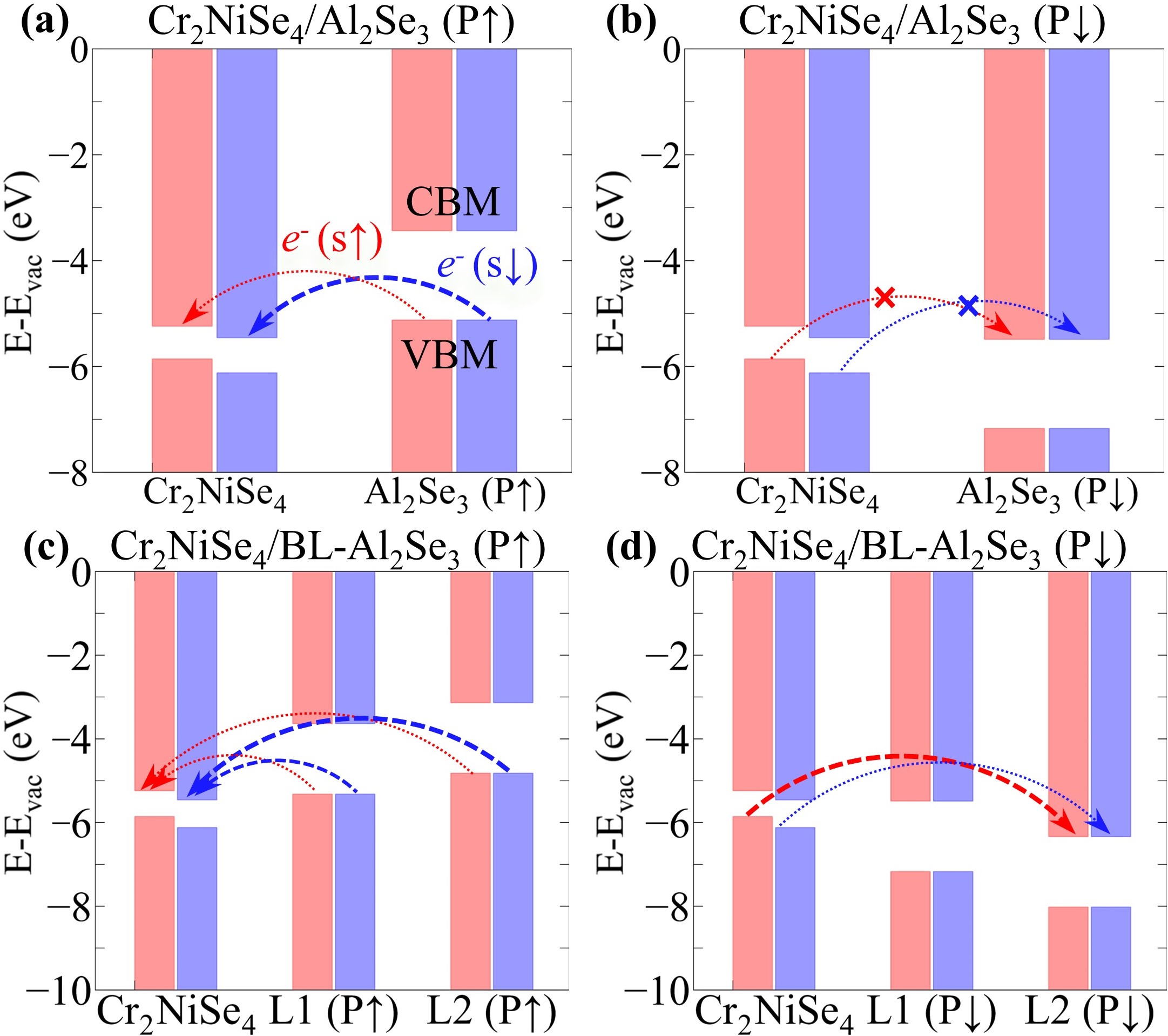}\\
		\caption{
			Band alignments in heterostructures.
			Band alignments respect to the vacuum level of Cr$_2$NiSe$_4$ at (a) Al$_2$Se$_3$ (P$\uparrow$), (b) Al$_2$Se$_3$ (P$\uparrow$), (c) bilayer Al$_2$Se$_3$ (P$\uparrow$) and (d) bilayer Al$_2$Se$_3$ (P$\uparrow$).
		}\label{Energy_level}
	\end{figure}
	
	The built-in potential from the ferroelectric layer directly dictates the relative alignment of the energy bands at the interface.
	Reversing the ferroelectric polarization causes a substantial shift in this alignment, which in turn governs the nature of the interfacial charge transfer, as depicted schematically in Figs. \ref{Energy_level}(a) and \ref{Energy_level}(b).
	For the Cr$_2$NiSe$_4$/Al$_2$Se$_3$(P$\uparrow$) heterostructure, the VBM of Al$_2$Se$_3$(P$\uparrow$) is higher than the CBM of Cr$_2$NiSe$_4$, resulting in a spontaneous injection of electrons from the Al$_2$Se$_3$ layer into the Cr$_2$NiSe$_4$ layer.
	Bader charge analysis quantifies this transfer, revealing that the Cr$_2$NiSe$_4$ layer gains a net charge of $2.66\times 10^{13} $ e$^-$/cm$^2$.
	This substantial electron doping raises the Fermi level of the system, causing it to cross the spin-down conduction band of Cr$_2$NiSe$_4$.
	Consequently, the material is transformed from a BMS into a HM state with spin-down polarized carriers, as confirmed by the layer-resolved density of states (DOS) shown in Fig. \ref{AC}(c).

	In contrast, for the Cr$_2$NiSe$_4$/Al$_2$Se$_3$(P$\downarrow$) heterostructure, the VBM of Cr$_2$NiSe$_4$ is lower than the CBM of Al$_2$Se$_3$, as shown in Fig. \ref{Energy_level}(b).
	This alignment creates a large energy barrier that largely suppresses interlayer charge injection.
	A much smaller charge transfer of $9.3\times 10^{12} $ e$^-$/cm$^2$ from Al$_2$Se$_3$ to Cr$_2$NiSe$_4$ is observed.
	As a result, the Cr$_2$NiSe$_4$ layer remains its BMS behavior, as shown by its DOS in Fig. \ref{AC}(f).

	Beyond the electronic properties, the ferroelectric state also strongly modulates the magnetic properties of the heterostructure, as listed in Tab. \ref{tab:FE}. 
	For the Cr$_2$NiSe$_4$/Al$_2$Se$_3$(P$\downarrow$) heterostructure, the magnetic coupling constants within the Cr$_2$NiSe$_4$ layer are nearly identical to those of the pristine monolayer, resulting in a robust $T\rm_C$ of 390 K.
	In contrast, the interlayer AFM coupling constants of the Cr$_2$NiSe$_4$/Al$_2$Se$_3$(P$\uparrow$) heterostructure significantly increase, resulting in stronger interlayer competition and a lower $T\rm_C$ of 152 K. 
	Further calculation details are available in the Supplementary Material \cite{SM}.

	
	In addition, we calculated the binding energy $E_b$ for Cr$_2$NiSe$_4$/X, via $E{\rm_b}=(E{\rm_{Cr_2NiSe_4/X}} - E{\rm_{Cr_2NiSe_4}} - E{\rm_{X}})/N{\rm_{tot}}$, where $E{\rm_{Cr_2NiSe_4/X}}$, $E{\rm_{Cr_2NiSe_4}}$, and $E{\rm_{X}}$ represent the total energies of heterostructure, monolayer Cr$_2$NiSe$_4$ and X, respectively, and $N_{tot}$ is the total atom number in the unitcell.
	The calculated results are -87.63 and -87.56 meV/atom for X as ML-Al$_2$Se$_3$(P$\uparrow$) and ML-Al$_2$Se$_3$(P$\downarrow$), respectively, indicating their stability.
	
	\begin{table*}[thb]
		\setlength{\tabcolsep}{1.5mm}
		\caption{
			Results of heterostructures.
			The band structure, binding energy $E\rm_b$, change of electron difference in three layers, coupling constants, MAE, magnetic ground state and $T\rm_{C}$ for Cr$_2$NiSe$_4$/Al$_2$Se$_3$(P$\uparrow$/P$\downarrow$) and Cr$_2$NiSe$_4$/BL-Al$_2$Se$_3$(P$\downarrow$/P$\downarrow$).
			The units of $E\rm_b$, $T\rm_C$ are meV/atom and K, respectively.
		}{\scalebox{1}{
				\begin{tabular}{c|c|c|c|c|c|c|c|c|c}
					\hline\hline
					\makebox[0.10 \textwidth][c]
					{Cr$_2$NiSe$_4$/X} &
					\multicolumn{9}{c}{Properties}
					\\
					\hline
					\makebox[0.12 \textwidth][c]
					{\multirow{2}{*}{X}} &
					\makebox[0.10 \textwidth][c]
					{\multirow{2}{*}{Carrier}}&
					\makebox[0.010 \textwidth][c]
					{$E\rm_b$}&
					\multicolumn{4}{c|}{Coupling constants (meV)} &
					\makebox[0.065 \textwidth][c]
					{MAE} &
					\makebox[0.065 \textwidth][c]{Ground}&
					\makebox[0.065 \textwidth][c]
					{\multirow{2}{*}{$T\rm_C$ (K)}}
					\\
					\cline{4-7}
					&
					&(meV/atom)
					&
					\makebox[0.055 \textwidth][c]{$J_\parallel S^2$}&
					\makebox[0.055 \textwidth][c]{$J_\parallel' S^2$}&
					\makebox[0.055 \textwidth][c]{$J_\perp S^2$}&
					\makebox[0.055 \textwidth][c]{$J_\perp' S^2$}&
					(meV)
					&State
					&
					\\
					\hline
					ML-Al$_2$Se$_3$ (P$\uparrow$)
					&e$^-$ (spin $\downarrow$ )
					&-87.63&-23.76  & 7.17&-31.28 & 10.03&0.17  & FM& 152\\
					ML-Al$_2$Se$_3$ (P$\downarrow$)&Semiconductor &-87.56&-25.94 & 5.62 &-36.25 &  1.25  &-0.02 & FM& 390 \\\hline
					BL-Al$_2$Se$_3$ (P$\uparrow$)&
					e$^-$ (spin $\downarrow$ )
					&-18.48
					&-27.19 & 5.62 & -36.25 &3.75 & 0.13 & FM& 305 \\
					BL-Al$_2$Se$_3$ (P$\downarrow$)&
					h$^+$ (spin $\uparrow$ ) 
					& -16.13
					&-19.69  & 14.37&-46.25 & -6.25 & 0.64 & FM& 400 \\
					\hline\hline
				\end{tabular}
		}}
		\label{tab:FE}
	\end{table*}

	\begin{figure*}[hbpt]
		\centering
		\includegraphics[width=1.6\columnwidth]{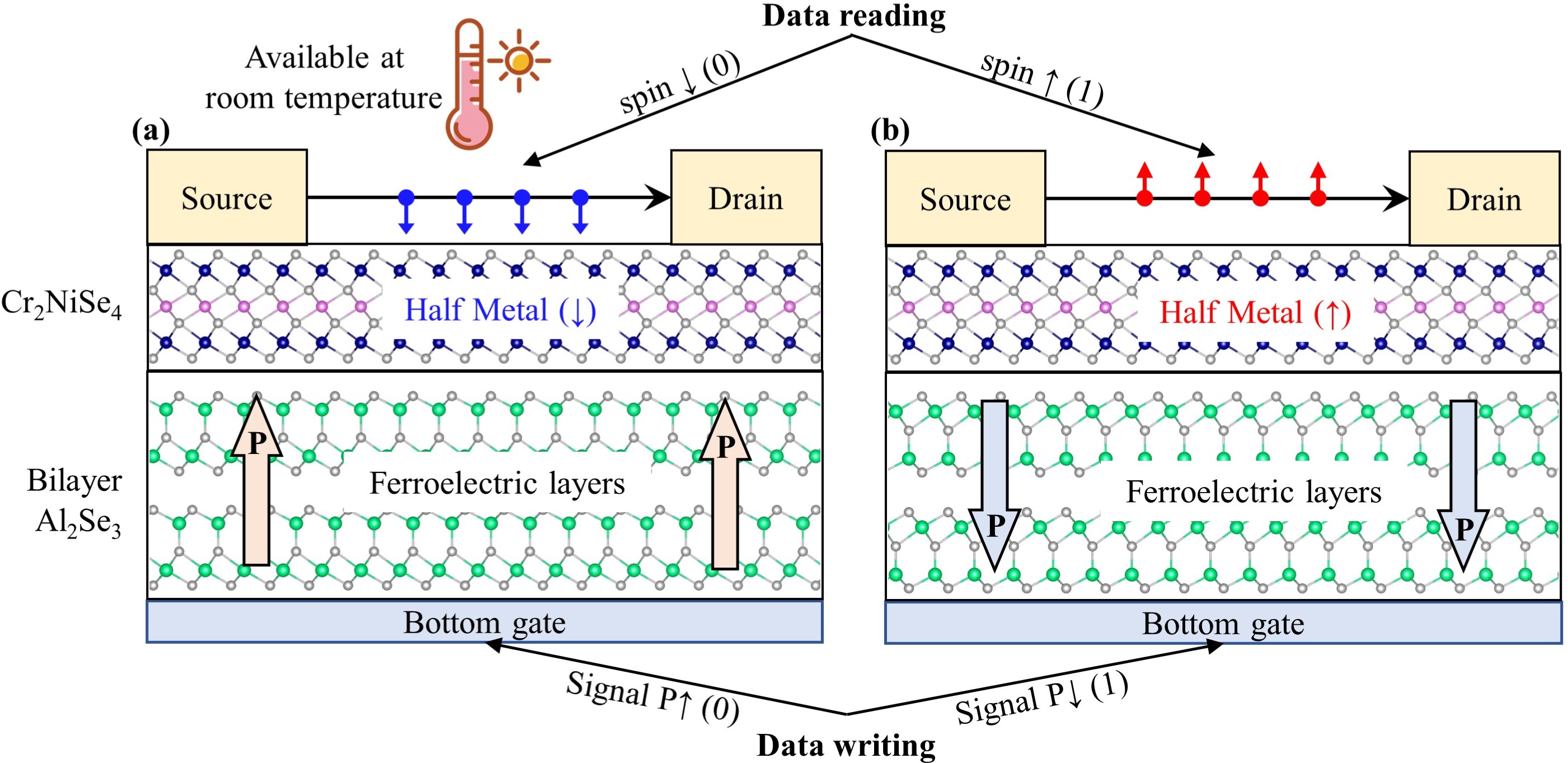}\\
		\caption{
			Multiferroic memories based on Cr$_2$NiSe$_4$/BL-Al$_2$Se$_3$.
			(a) Cr$_2$NiSe$_4$ at BL-Al$_2$Se$_3$($\uparrow$) , chosen as state '0'.
			(b) Cr$_2$NiSe$_4$ at BL-Al$_2$Se$_3$($\uparrow$) chosen as state '1'.
			The data writing is controlled by switching the ferroelectric polarization of BL-Al$_2$Se$_3$ with electric single.
			Data reading is based on the reading carriers' spin polarization of Cr$_2$NiSe$_4$.
		}\label{Device}
	\end{figure*}
	
	\subsection{Cr$_2$NiSe$_4$ at ferroelectric gate bilayer Al$_2$Se$_3$}

	To further modulate Cr$_2$NiSe$_4$, we constructed the heterostructure Cr$_2$NiSe$_4$ with bilayer Al$_2$Se$_3$ (BL-Al$_2$Se$_3$).
	Their structure are shown in Figs. \ref{A2C}(a) and \ref{A2C}(d), respectively.
	Different stack models are discussed in Supplementary Material \cite{SM}.
	We label the Al$_2$Se$_3$ layers as L1 (closer to Cr$_2$NiSe$_4$) and L2 (further away).
	The presence of two ferroelectric layers creates a substantially larger electrostatic potential drop across the heterostructure compared to the monolayer case, as shown in Figs. \ref{A2C}(b) and \ref{A2C}(e).

	This enhanced electrostatic potential enables more powerful tuning of the BMS.
	In Cr$_2$NiSe$_4$/BL-Al$_2$Se$_3$(P$\uparrow$) heterostructure, the VBMs of both Al$_2$Se$_3$ layers are higher than the CBM of Cr$_2$NiSe$_4$, as shown in Fig. \ref{Energy_level}(c), leading to electron injection from both Al$_2$Se$_3$ layers into Cr$_2$NiSe$_4$. 
	Bader charge analysis confirms a total charge transfer of 3.95 $\times$10$^{13}$ electrons/cm$^2$.
	This electron doping raises the Fermi level into the spin-down conduction band, transforming the Cr$_2$NiSe$_4$ into a spin-down HM, as confirmed by the layer-resolved DOS in Fig. \ref{A2C}(c).

	In Cr$_2$NiSe$_4$/BL-Al$_2$Se$_3$(P$\downarrow$), similar to Cr$_2$NiSe$_4$/Al$_2$Se$_3$(P$\downarrow$), the VBM of Cr$_2$NiSe$_4$ is lower than the CBM of L1, preventing interlayer charge injection between L1 and Cr$_2$NiSe$_4$.
	However, the large potential difference between the L1 and L2 layers (Fig. \ref{Energy_level}(d)) creates a strong driving force for charge redistribution.
	Due to the polarization, the CBM of L2 is lower than the VBM of Cr$_2$NiSe$_4$ (Fig. \ref{Energy_level}(d)), leading to an indirect electron injection from Cr$_2$NiSe$_4$ to L2, mediated by L1.
	Bader charge analysis reveals that all electrons transferred from Cr$_2$NiSe$_4$ are accumulated in layer L2, with a total charge transfer of 1.0$\times 10^{12}$ cm$^{-2}$.
	This charge transfer lowers the Fermi level into the spin-up valence band, transforming the Cr$_2$NiSe$_4$ into a half-metal with fully spin-up polarized carriers, as shown in the layer-resolved DOS in Fig. \ref{A2C}(f).
	
	Furthermore, we investigated the magnetic properties and structural stability of these bilayer heterostructures.
	The key results are summarized in Table \ref{tab:FE} \cite{SM}.
	Both the P$\uparrow$ and P$\downarrow$ configurations are found to retain high $T\rm_C$ above room temperature, along with in-plane magnetic anisotropy.
	To confirm their structural viability, we calculated the binding energies for the interfaces.
	The results, -18.48 meV/atom for the P$\uparrow$ state and -16.13 meV/atom for the P$\downarrow$ state, are both negative, indicating that the heterostructures are stable.
	Thus, the bilayer Al$_2$Se$_3$ gate provides a robust platform for achieving nonvolatile control over the carrier spin polarization in Cr$_2$NiSe$_4$.

    \subsection{Devices based on Cr$_2$NiSe$_4$/BL-Al$_2$Se$_3$ heterostructure}
    
    The information age has driven the development of smaller, faster, low-energy, room-temperature, and tunable spintronic devices as a central focus of current research.
    2D multiferroic materials are ideal platforms for next-generation data storage due to their unique combination of ferroelectric polarization and magnetism, enabling energy-efficient voltage control of magnetism.
    These characteristics make them particularly attractive for future device applications \cite{Yang2023a,Wang2021,Zhao2019,Hu2015,Zhao2019,Kang2024,Tao2025}.

    Based on our findings, the Cr$_2$NiSe$_4$/BL-Al$_2$Se$_3$ heterostructure provides an ideal platform for designing nanoscale spintronic devices, such as a multiferroic nonvolatile memory.
    The conceptual design of this 2D device, illustrated in Fig. \ref{Device}, leverages the ability to non-volatilely switch the Cr$_2$NiSe$_4$ channel between two distinct and oppositely spin-polarized half-metallic states, which serve as the binary '0' and '1' for data storage.
    The device operates on a clear "write-read" principle based on magnetoelectric coupling.
    The 'write' operation is performed by applying a transient electrical pulse across the bilayer Al$_2$Se$_3$ gate, which deterministically switches its ferroelectric polarization and, in turn, sets the Cr$_2$NiSe$_4$ into either the spin-down (representing '0') or spin-up (representing '1') half-metallic state.
    The 'read' operation is accomplished by sensing the spin-dependent electrical conductance of the Cr$_2$NiSe$_4$ channel via current or light detectors.

    The proposed multiferroic memory shows significant advantages.
    First and foremost, its operation is inherently non-volatile.
    Using a ferroelectric material to maintain the memory state, the device eliminates the need for standby power, drastically reducing overall energy consumption.
    The write operation itself is also highly efficient, achieved with a transient electrical signal to switch the polarization.
    Furthermore, the device features a non-destructive readout mechanism without disturbing the underlying ferroelectric polarization.
    The atomically thin structure of the heterostructure facilitates aggressive downscaling of the device footprint, making it an ideal candidate for future high-density memory integration.
    Finally, the high intrinsic $T\rm_C$ of the Cr$_2$NiSe$_4$ channel guarantees robust and reliable operation at room temperature, fulfilling a fundamental prerequisite for any practical application.
    Collectively, these features demonstrate that our design provides a concrete and promising pathway toward high-performance, nonvolatile memory based on a novel 2D multiferroic system.

	\section{Conclusion}
	Based on the DFT calculations, we predict some room temperature magnetic semiconductors by intercalation in 2D magnetic materials realized in experiments.
	Their stability were indicated by formation energy, phonon spectrum and molecular dynamics simulations.
	Specifically, a 2D ferromagnetic BMS, Cr$_2$NiSe$_4$, was obtained by intercalating Ni atoms into the van der Waals gap of bilayer CrSe$_2$, exhibiting a band gap of 0.40 eV and a $T\rm_C$ of 352 K. 
	To manipulate the properties of Cr$_2$NiSe$_4$, we constructed heterostructures by coupling Cr$_2$NiSe$_4$ with monolayer and bilayer ferroelectric Al$_2$Se$_3$. 
	Different ferroelectric polarization directions result in distinct band alignments within the heterostructure, leading to electron or hole injection into Cr$_2$NiSe$_4$ and altering its properties.
	Nonvolatile spin polarization control of Cr$_2$NiSe$_4$ was achieved by switching the ferroelectric polarization.
	The heterostructures Cr$_2$NiSe$_4$/ML-Al$_2$Se$_3$(P$\uparrow$ or P$\downarrow$) and Cr$_2$NiSe$_4$/BL-Al$_2$Se$_3$(P$\downarrow$ or P$\downarrow$) exhibit HM(spin $\downarrow$) or BMS, HM(spin $\downarrow$ or $\uparrow$) properties, respectively.
	Based on Cr$_2$NiSe$_4$/BL-Al$_2$Se$_3$, multiferroic nonvolatile memories are proposed, where data writing is controlled by the ferroelectric polarization of BL-Al$_2$Se$_3$, and data reading is based on the distinct spin polarizations of carriers in Cr$_2$NiSe$_4$.
	This device offers the advantages of convenient nondestructive reading and writing, high-temperature operation, and low energy consumption for data storage.
	Our work not only reports a 2D BMS with $T\rm_C$ above room temperature, but also present a feasible strategy to achieve nonvolatile electrical control of 2D BMS.

	\section{Method}
	All calculations were based on the DFT as implemented in the Vienna ab initio simulation package (VASP) \cite{Kresse1996}.
	The exchange-correlation potential is described by the PBE form with the generalized gradient approximation (GGA) \cite{Perdew1996}.
	The electron-ion potential is described by the projector-augmented wave (PAW) method \cite{Bloechl1994}.
	We carried out the calculation of PBE + $U$ with $U$ = 4 eV for 3d electrons.
	To prevent interlayer interaction in the supercell of 2D systems, the vacuum layer of 20~\AA~is included.
	The plane-wave cutoff energy is set to be 650 eV.
	The 9$\times$9$\times$1 $\Gamma$ center K-points was used for the Brillouin zone (BZ) sampling.
	To obtain accurate results of magnetic anisotropy energy (MAE), K-points were chosen as $\Gamma$-centered 18$\times$18$\times$1.
	The structures of all materials were fully relaxed, where the convergence precision of energy and force was $10^{-6}$ eV and $-10^{-2}$ eV/\AA, respectively.
	Phonon spectra were calculated through the PHONOPY package \cite{Togo2015}.
	The Phonon spectra of monolayer Cr$_2$NiSe$_4$ and bulk Cr$_2$NaSe$_4$ were calculated with $5\times5\times1$ and $3\times3\times3$ supercells with 175 and 108 atoms, respectively.
	The van der Waals interactions were considered by using the DFT-D3 method \cite{Grimme2006}.
	The molecular dynamics simulations were performed at 300 K, using the NVT ensemble (constant temperature and volume) and running for 6 ps.
	4$\times$4$\times$1 supercells of predicted monolayers containing 112 atoms were considered.
	
	The Heisenberg type Monte Carlo simulations were performed to investigate the magnetic properties of bilayer and bulk CrX$_2$ (X = S, Se) intercalated with different atoms. 
	For bilayer CrX$_2$ intercalated with magnetic atoms (e.g., monolayer Cr$_2$NiSe$_4$), a 40$\times$40$\times$1 supercell containing 4800 magnetic sites was used. 
	For bilayer CrX$_2$ intercalated with nonmagnetic atoms (e.g., monolayer Cr$_2$CdSe$_4$), a 45$\times$45$\times$1 supercell containing 4050 magnetic sites was used. 
	For bulk CrX$_2$ intercalated with nonmagnetic atoms (e.g., bulk CrKTe$_2$), a 16$\times$16$\times$16 supercell containing 4096 magnetic sites was employed. 
	For each temperature, 3$\times$10$^5$ Monte Carlo steps were carried out, with the last one-third used for calculating temperature-dependent physical quantities.

    \section {Acknowledgements}
    This work is supported by National Key R\&D Program of China (Grant No. 2022YFA1405100), Chinese Academy of Sciences Project for Young Scientists in Basic Research (Grant No. YSBR-030), and Basic Research Program of the Chinese Academy of Sciences Based on Major Scientific Infrastructures (Grant No. JZHKYPT-2021-08).

    %

\end{document}